\documentclass[aps,prl,reprint,superscriptaddress]{revtex4-1}

\usepackage{xcolor}
\usepackage{tikz}

\usepackage{amssymb,amsmath}
\usepackage{graphicx}
\usepackage{upgreek}
\usepackage[english]{babel}
\usepackage[T1]{fontenc}
\usepackage[utf8]{inputenc}
\usepackage{graphicx}
\usepackage{bm}
\def\pmb#1{\setbox0=\hbox{#1}
\kern-.025em\copy0\kern-\wd0 \kern-.05em\copy0\kern-\wd0
\kern-.025em\raise.0433em\box0}

\begin{document}
%\doublespace

\title{Cloaking in-plane elastic waves with swiss rolls}

\author{Younes Achaoui}
\address{Institut FEMTO-ST, CNRS, Universite de Bourgogne Franche-Comte, 25044 Besancon Cedex, France}
\author{Andr\'e Diatta}
\address{Aix Marseille Univ, CNRS, Centrale Marseille, Institut Fresnel,
%\\Campus universitaire de Saint-J\'er\^ome,
13013 Marseille, France}
\author{Muamer Kadic}
\address{Institut FEMTO-ST, CNRS, Universite de Bourgogne Franche-Comte, 25044 Besancon Cedex, France \\
Institute of Nanotechnology, Karlsruhe Institute of Technology (KIT), 76128 Karlsruhe, Germany}
%\author{Martin Wegener}
%\address{Institute of Nanotechnology, Karlsruhe Institute of Technology (KIT), 76128 Karlsruhe, Germany \\
%Institute of Applied Physics, Karlsruhe Institute of Technology (KIT), 76128 Karlsruhe, Germany
%}
\author{S\'ebastien Guenneau}
\address{Aix Marseille Univ, CNRS, Centrale Marseille, Institut Fresnel,
%\\Campus universitaire de Saint-J\'er\^ome,
13013 Marseille, France}
\email{sebastien.guenneau@fresnel.fr}

\begin{abstract}
We propose a design of cylindrical elastic cloak for coupled in-plane shear waves consisting of concentric layers of sub-wavelength resonant stress-free inclusions shaped as swiss-rolls.
The scaling factor between inclusions' sizes is according to Pendry's transform. Unlike the hitherto known situations, the present geometric transform starts from a Willis medium and further assumes that displacement fields ${\bf u}$ in original medium and ${\bf u}'$ in transformed medium remain unaffected (${\bf u}'={\bf u}$), and this breaks the minor-symmetries of the rank-4 and rank-3 tensors in the Willis equation that describes the transformed effective medium. We achieve some cloaking for a shear polarized source at specific, resonant sub-wavelength, frequencies, when it is located near a clamped obstacle surrounded by the structured cloak.
%We interpret our results in terms of an effective elastic medium described by a modified Willis equation i.e. with the usual rank 3 and 2 tensors, but with a rank-4 elasticity tensor without the minor symmetry.
Such an effective medium allows for strong Willis coupling [Quan et al., Physical Review Letters {\bf 120}(25), 254301 (2018)], notwithstanding potential chiral elastic effects [Frenzel et al., Science {\bf 358}(6366), 1072 (2017)], and thus mitigates roles of Willis and Cosserat media in the achieved elastodynamic cloaking. 

\end{abstract}

\maketitle

\section{Introduction}

Following the paper of Milton, Briane and Willis  \cite{Milton}, a new field has emerged in metamaterials, that of transformed elastic media enabling to make a region neutral to fully coupled cylindrical \cite{apl2009} and spherical \cite{diatta-apl2014} elastic waves. There are different routes to elastic cloaking which have been listed in \cite{Norris,Mccall}, and we shall focus here on one of these based on the concept of unconventional effective dynamic properties enabling both minor symmetry breaking in the rank-4 elasticity tensor as well as non vanishing rank-3 and 2 tensors in the Willis model \cite{Willis} near resonant frequencies of certain types of stress-free inclusions shaped as swiss-rolls. Swiss-rolls were introduced in the context of electromagnetic metamaterials for artificial chirality \cite{pendry}, and such magneto-optic coupling recently found a counterpart in elasticity \cite{Frenzel,fernandez2019,kadic2019,kadic2019b}.

We note that the idea of Willis media \cite{Willis} described by
\begin{equation}
\nabla_{\bf x}\cdot\left({\bf C}: \nabla_{\bf x}{\bf u}+{\bf S}\cdot{\bf u}\right)+\underline{\underline{\rho}}\omega^2 {\bf u} = -{\bf D}: \nabla_{\bf x}{\bf u}\; ,
\label{willis1}
\end{equation}
with the rank-4 elasticity tensor ${\bf C}$  having all its minor and major symmetries, as well as the rank-3 elasticity tensors ${\bf S}$ and ${\bf D}$ such that $D_{pqr}=-S_{qrp}$, and the rank-2  (symmetric) density tensor $\underline{\underline{\rho}}$, was introduced as a promising route to elastodynamic cloaking and as a solution to the non-invariance of the Navier equation under general change of coordinates \cite{Milton}. This was done thanks to a properly chosen gauge linking the displacement fields ${\bf u}$ and ${\bf u}'$ through the Jacobian of the transformation. As pointed out in \cite{apl2009,diatta-apl2014}, if one assumes that ${\bf u}={\bf u}'$ the Navier equation
\begin{equation}
\nabla_{\bf x}\cdot\left({\bf C} : \nabla_{\bf x}{\bf u}\right)+\rho\omega^2 {\bf u} = {\bf 0} \; ,
\label{navierhetero}
\end{equation}
retains its form under coordinate change, but the elasticity tensor ${\bf C}$ loses its minor symmetry. Other choices of the gauge lead to different types of transformed media \cite{Norris}.

In this letter, we stress that we start from a Willis material, as our background material and transform it into a new material with some specific properties. Namely , if we consider the coordinate change $\phi : {\bf x}=(x_1,x_2,...)\longmapsto {\bf x'}=(x_1'({\bf x}),x_2'({\bf x}),...)$ and we impose that the displacement ${\bf u}={\bf u}'$ in Willis's equation (\ref{willis1}),  this equation is actually form invariant, but the tensors therein lose their minor symmetries. This kind of transformed medium therefore has in it some features of Cosserat media and the expression of transformed tensors ${\bf C}'$, ${\bf D}'$, ${\bf S}'$ and $\underline{\underline{\rho}}'$ that reflect the minor symmetry breaking is given in Appendix A.

Our observation opens interesting avenues for the design of cylindrical elastodynamic cloaks via homogenization approaches combining recent findings in metamaterials displaying strong Willis coupling \cite{Quan,Melnikov} and chiral elasticity features \cite{Frenzel,kadic2019b}, as we shall see in the sequel. To exemplify the usefulness of the transformed Willis equation with non-fully symmetric elasticity tensors, we propose to design a microstructured cloak consisting of swiss-rolls displaying the usual features encountered in both Cosserat and Willis media in the low frequency limit. We consider a simplified form of Navier equation that governs the propagation of elastic
waves in an isotropic homogeneous elastic medium
\begin{equation}
(\lambda+2\mu)\nabla\nabla\cdot{\bf u} - \mu \nabla\times\nabla\times {\bf u} + \rho \frac{\partial^2}{\partial t^2} {\bf u} = {\bf 0}
\label{navierhomo}
\end{equation}
where $\lambda$ and $\mu$ are the compressional and shear Lam\'e coefficients and $\rho$ is
the density.

If there are inclusions in the homogeneous medium, one can supply (\ref{navierhomo}) with boundary conditions, such as clamped ${\bf u}={\bf 0}$
or stress-free $\sigma({\bf u})\cdot{\bf n}=({\bf C}:\epsilon({\bf u}))\cdot {\bf n}={\bf 0}$
where ${\bf C}$ is the rank-4 elasticity tensor with entries $C_{ijkl}=\lambda\delta_{ij}\delta_{kl}+\mu(\delta_{ik}\delta_{jl}+\delta_{il}\delta_{jk})$
and $\sigma({\bf u})$ and $\epsilon({\bf u})$ are the rank-2 stress and strain tensors with entries $\sigma_{ij}=\lambda\varepsilon_{kk}\delta_{ij}+2\mu\varepsilon_{ij}$
and $\varepsilon_{ij}=1/2(\partial u_i/\partial x_j+\partial u_j/\partial x_i)$, respectively, ${\bf n}$ being the
outward pointing normal to the boundary of inclusions. It is then easily seen from the Helmholtz decomposition
$
{\bf u} = \nabla\Phi + \nabla\times {\bf\Psi}
\; , \; \nabla\cdot {\bf\Psi} = 0 \; ,
%\label{potential}
$
with scalar (pressure related) and vector (shear related) Lam\'e potentials
$\Phi$ and ${\bf\Psi}$ that for stress-free inclusion, pressure (p) and shear (s) waves in (\ref{navierhomo}) are now coupled. Indeed, there is a conversion of p in s waves (and vice versa) at any stress-free boundary,
and this coupling has been used previously notably for opto-elastic switches in arrays of stress-free holes in silica \cite{Russell}. 

\begin{figure*}[h!]
\centering{
\includegraphics[scale=1]{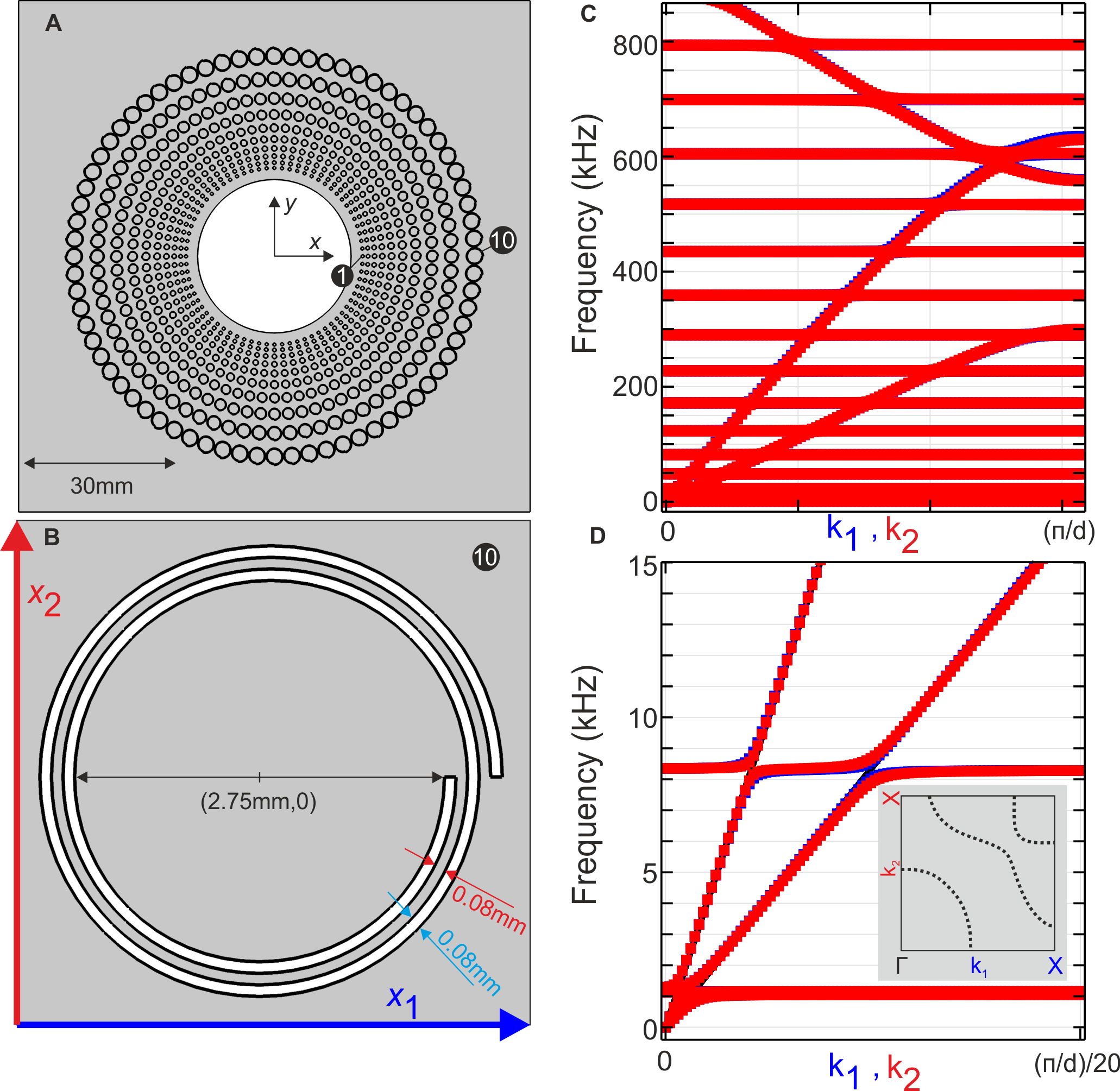}
\caption{Geometrical characteristics and dispersion properties of the investigated model. (A) Geometry of the entire cloak; (B) Zoom on an elementary cell; (C) Band diagram for a Bloch vector ${\bf k}$ running along $\Gamma M$ (${\bf k}=(k_1,0)$ with $k_1\in [0,\pi/d]$) and along $MX$ (${\bf k}=(\pi/d,k_2$) with $k_2\in [0,\pi/d]$), showing the effective medium is isotropic; (D) Zoom-in in the neighborhood of $\Gamma$, where one notes the avoided crossings at resonances around $1$ kHz and $8$ kHz;. These dispersion curves serve as a guide for our homogenized model with an inset showing the isofrequencies around the resonance (approximation of a Willis-type medium).}
\label{fig1}}
\end{figure*}

Let us now assume that a homogeneous medium is structured with a square array of stress-free inclusions shaped as swiss-rolls invariant along the $x_3$-axis.
%Since we assume that the swiss-rolls are invariant
%along the $x_3$-axis,
Thanks to this invariance, we can consider in-plane
coupled  shear and pressure elastic waves on one hand,
with unknown $(u_1,u_2,0)$ and anti-plane shear
waves with unknown $(0,0,u_3)$ in (\ref{navierhomo}),
on the other hand. We focus on the former.
The periodicity of the
cladding implies that the in-plane displacement field ${\bf u}=(u_1,u_2)$ satisfies the Floquet-Bloch theorem:
\begin{equation}
{\bf u}(x_1+d,x_2+d)={\bf u}_{\bf k}(x_1,x_2) \exp(i(k_1d+k_2d))
\label{floquetbloch}
\end{equation}
where ${\bf k}=(k_1,k_2)$ is the Bloch vector which describes the first Brillouin
zone (BZ) $\Gamma$MX in the reciprocal space, with $\Gamma=(0,0)$, $M=(\pi/d,0)$
and $X=(\pi/d,\pi/d)$ and $d$ the array pitch. One can then look for eigenfrequencies
$\omega_{\bf k}$ and associated Floquet-Bloch eigenfields ${\bf u}_{\bf k}$ solutions of (\ref{navierhomo}),
and by letting ${\bf k}$ vary within IBZ we compute some dispersion diagrams.
We display
in Fig. \ref{fig1} geometric characteristics of the swiss-rolls under study (panels A, B) and
associated dispersion curves along $\Gamma$M (green curves) and $MX$ (red curves), see panel C. One notes that flat
bands correspond to localized modes associated with resonances of the swiss-rolls. It is
also observed in panel D that wavespeed of s waves differs along $\Gamma$M and $MX$ directions, which
is interpreted as a dynamic anisotropic mass density.
%We shall come back to this in the sequel.
Properties of the effective symmetric rank-4 tensor ${\bf C}$ and rank-3 elasticity tensors ${\bf S}$ and ${\bf D}$ such that $D_{pqr}=\textcolor{red}{-}S_{qrp}$ and the rank-2  density tensor $\underline{\underline{\rho}}$ are inferred from a retrieval method such as what was done in \cite{Frenzel,kadic2019b}, or alternatively from a direct Bloch-wave \cite{Nassar} homogenization approach
applied to the doubly periodic array of identical swiss-rolls in Cartesian coordinates, see Fig. \ref{fig1}.

%, make it difficult to achieve
%the required elasticity parameters via a classical homogenization approach and so one needs to apply so-called Bloch-wave homogenization schemes like in \cite{Andrianov, Auriault,Nassar}, which leads to
%\begin{equation}
%(\nabla_{\bf X}+i{\bf k})\cdot\left({\bf C} : [(\nabla_{\bf X}+i{\bf k})\otimes^s{\bf u}_{\bf k}]\right)+\rho\omega^2 {\bf u}_{\bf k} = {\bf 0} \; ,
%\label{navierheterobloch}
%\end{equation}
%where $\otimes^s$ denotes tensor product with symmetrization.
%The strain field is expressed as $\epsilon_{\bf k}=(\nabla+i{\bf k})\otimes^s{\bf u}_{\bf k}$.

However, as it has been explained above, transformation physics affects the Willis equation (\ref{willis1}), although it retains its form if we assume that ${\bf u}={\bf u}'$, and so when we apply Pendry's transform to the doubly periodic array of swiss-rolls, the symmetry of the tensors in the effective Willis equation gets broken, and besides from that they become spatially varying. Therefore, the effective Willis equation describing the cloak with gradually varying swiss-rolls in Fig. \ref{fig2} has the form of Eq. (7)-(10) in Appendix A.

So when we map the doubly periodic array of swiss-rolls on a transformed medium using Pendy's transform, the transformed Willis medium now has the Cosserat features built in it. The swiss-roll based cloak is an example that illustrates this type of combined mechanisms, in which cloaking is due to both Willis and Cosserat materials. Indeed, a wave is usually characterized by its polarization, a direction of the wavenumber, a frequency and a rated velocity. The dynamic density can straightforwardly be omitted since we are exciting the propagation at a unique nominal frequency. However, the inertial behavior of the swiss-rolls entails a direction change of the wave propagation to circumvent the obstacle and is naturally accompanied by modes conversion (each inclusion becomes a secondary source of waves) . Mathematically speaking, this involves both the symmetry breaking of the elastic tensor and third order tensor of the Willis-type equation. Our futur goal is to rigorously quantify the weighting of each contribution with regard to geometrical and physical properties of the swiss-rolls. Interestingly, similar effective parameters for a chiral Willis medium have been deduced from a retrieval method in \cite{Kadic} applied to the mechanical metamaterial first introduced in \cite{Frenzel,kadic2019b} in the context Eringen equations which are the counterpart of bianisotropic equations in optics. The magneto-optic coupling is actually easily seen using classical homogenization techniques in \cite{physicab}, and same techniques could be applied to the effective medium description of the Willis coupling for our array of swiss-rolls.
%\textcolor{green}{A virer: This ambitious purpose is not under the scope of the present paper since it needs further mathematical development and experimental validation.}

{However, one can alternatively deduce these features from the reading of band diagrams. When a bunch of resonant elements meet the wave propagation, a strong coupling between the so-called continuum and the resonators may occur. This can directly be identified in the band diagrams in Fig. 1 through band repealing between a polarized continuum and the flat mode describing the energy trapping in the resonator. This level repulsion can reach its maximum with the appearance of band gaps. The latter describes the energy prohibition inside the periodic structure through a total reflection, energy storage or conversion to other types of modes. A straight crossing between the bands reveals no interaction between the resonators and the continuum as it has been reported in \cite{Achaoui}. In the latter paper, we have evaluated the potentiality of a resonator to drastically change the direction of the wave for focusing purposes. In Fig. \ref{fig1}-A, we show a sketch of the swiss-roll based cloak with a zoom inset in Fig. \ref{fig1}-B. In Fig. \ref{fig1}-C, we depict the normalized band structure %(\textcolor{green}{il va falloir le mettre en normalisé pour ne pas parler de taille de cellule elementaire, surtout que nous avons une geometrie qui fait varier la cellule elementaire})
of a periodic structure made of inclusions shaped in swiss-roll resonators (Fig. \ref{fig1}-B). This band diagram shows mainly the two modes longitudinal and transverse starting from $\Gamma$ point and tremendous flat bands describing the resonance frequencies of the inclusion. It is worth noting that the number and the position of these bands in a determined range of frequency depend directly on the length of the spiral constituting the swiss-rolls. Hence, the cloak has been conceived in a way that most of the resonances are gathered in a tiny range of frequency. This choice was made to optimize the functionality of the inclusions while rolling. A zoom of the band structure near a resonance frequency is illustrated in Fig. \ref{fig1}-D. We can clearly observe that the flat band and the continuum repeal slightly from each other without creating a bandgap. Though the inclusion shaped as a swiss-roll is a bad candidate to achieve perfect reflectors, at this stage we are confident that this weak coupling to the continuum added to the potential of the inclusion to rotate under an incoming wave would contribute drastically to deflect the wave propagation. Furthermore, the level repulsion band anti-crossing between the flat mode and the continuum depends on the direction of the propagation just as well as the inclusion orientation (Fig. \ref{fig1}-D).
%(ce serait bien qu'on zoom davantage sur le anti-crossing pour montrer la difference entre (ox) et (oy) comme une ancienne version de Muamer).
In order to illustrate more this more or less strong coupling, we computed the isofrequency contours. In the inset of Fig. \ref{fig1}-D, the latter were evaluated around a frequency resonance. To be more consistent, let's split the Irreducible Brillouin Zone into to subsurfaces; ie GXM and GYM. Three bands (P, S and coupled PS) are identified around the frequency 8 kHz and each one is extended barely the same way in the two subsurfaces. If we look more closely we can notice that two kinds of anisotropies can be observed. The first one is the position of the wavenumbers. We can remark that for both polarizations P and S, the wave velocity toward GX is slightly fast compared to GY. The second anisotropy concerns the wave trapping (or coupling between the continuum and the resonator). We stress here that this coupling depends not on the wavevector but on the polarization of the wave (note the line width of the curves). } 

We test our cloak in Fig. 2 near resonances of the swiss-rolls, which have been scaled up and down with respect to Fig. 1, depending upon whether they are located on outer or inner, rings of the cloak in Fig. 2. We consider the frequency range from $9.6$ to $9.9$ kHz and pick up some resonant frequencies of some swiss-rolls. Upon inspection of the case of a shear-polarized point source in homogeneous medium (first row), same source in presence of a clamped obstacle without cloak (second row), with cloak (third row) and with a cloak without the proper design (fourth row), we deduce that cloaking is achieved i.e. the magnitude of the shear wave is recovered in forward scattering in third column, although with a slight phase delay induced by the longer wave trajectory induced by the cloak design. To exemplify the mechanism of the cloak, we show a magnified view of these plots in Fig. 3.

In this letter, we have proposed to approximate a Willis-type elastodynamic cloak with an elastic isotropic medium structured with stress-free swiss-rolls.
We consider a coordinate change $\phi$ such that ${\bf u}'({\bf x})={\bf u}({\bf x})$, in which case the transformed Willis equation has the
exact same structure as (\ref{willis1}) , but with a transformed elasticity tensor ${\bf C}'$ without the minor symmetries and same for the rank-3 tensors. Note however, that the density could be a scalar, and in any case it is fully symmetric. The cloak we have designed is thus neither totally of the Willis type \cite{Milton}, nor totally of the Cosserat type %\underline{\underline{\rho}}'$
\cite{apl2009, diatta-apl2014}. Finally, we note the alternative route of direct lattice transforms \cite{buckmann15,kadic2019c,nassar18} towards elastodynamic cloaking, which does not make use of resonant structural elements and thus follows a different protocol. In the near future, we would like to compare numerically and experimentally the efficiency of our cloak's design with those in \cite{kadic2019c,nassar18} in various scenarios. 

\begin{figure*}[h!]
\centering{
\includegraphics[width=150mm]{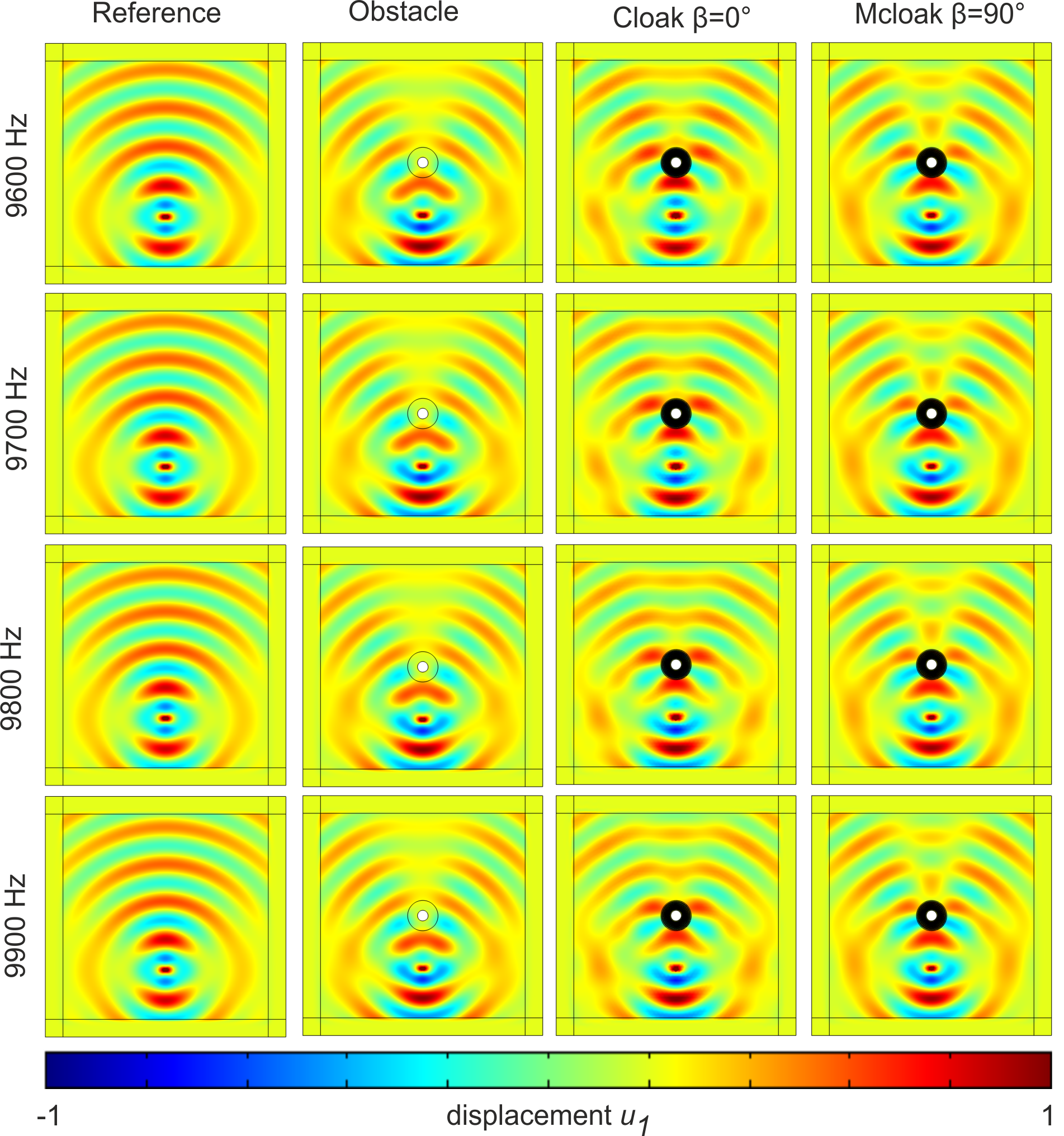}
\caption{ In-plane shear elastic wave generated by a point force located at $(x_1,x_2)=(0,-150)$ and oriented along the $x-axis$. This S wave propagates within an isotropic homogeneous elastic bulk (here PMMA) with a cloak centered at (0,0) of inner radius $r_1=1.5$ cm and outer radius $r_2=4$ cm and consisting of 11 concentric layers of swiss-rolls made of a soft material ($\lambda = 6.10^5 Pa$ and $\mu = 4.10^4$ Pa). The wave frequency ranges from $9.6$ to $9.9$ kHz.
%(what corresponds to a wavelength of $9.3 cm$).
Note that Cartesian elastic Perfectly Matched Layers have been set on either sides of the square computational domain. First column is for the shear-polarized point source in PMMA (benchmark); Second column has a clamped obstacle centered at (0,0) of radius $r_0=3$ cm; Third column is for the source with clamped obstacle and cloak.
%(d) and (e) are close up views showing resonances within the swiss-rolls responsible for the chiral effect.
Fourth column is same when the swiss-rolls have been tilted through an angle $\beta=\pi/2$ about their gravity center.}
\label{fig2}}
\end{figure*}

\begin{figure*}[h!]
\centering{
\includegraphics[width=150mm]{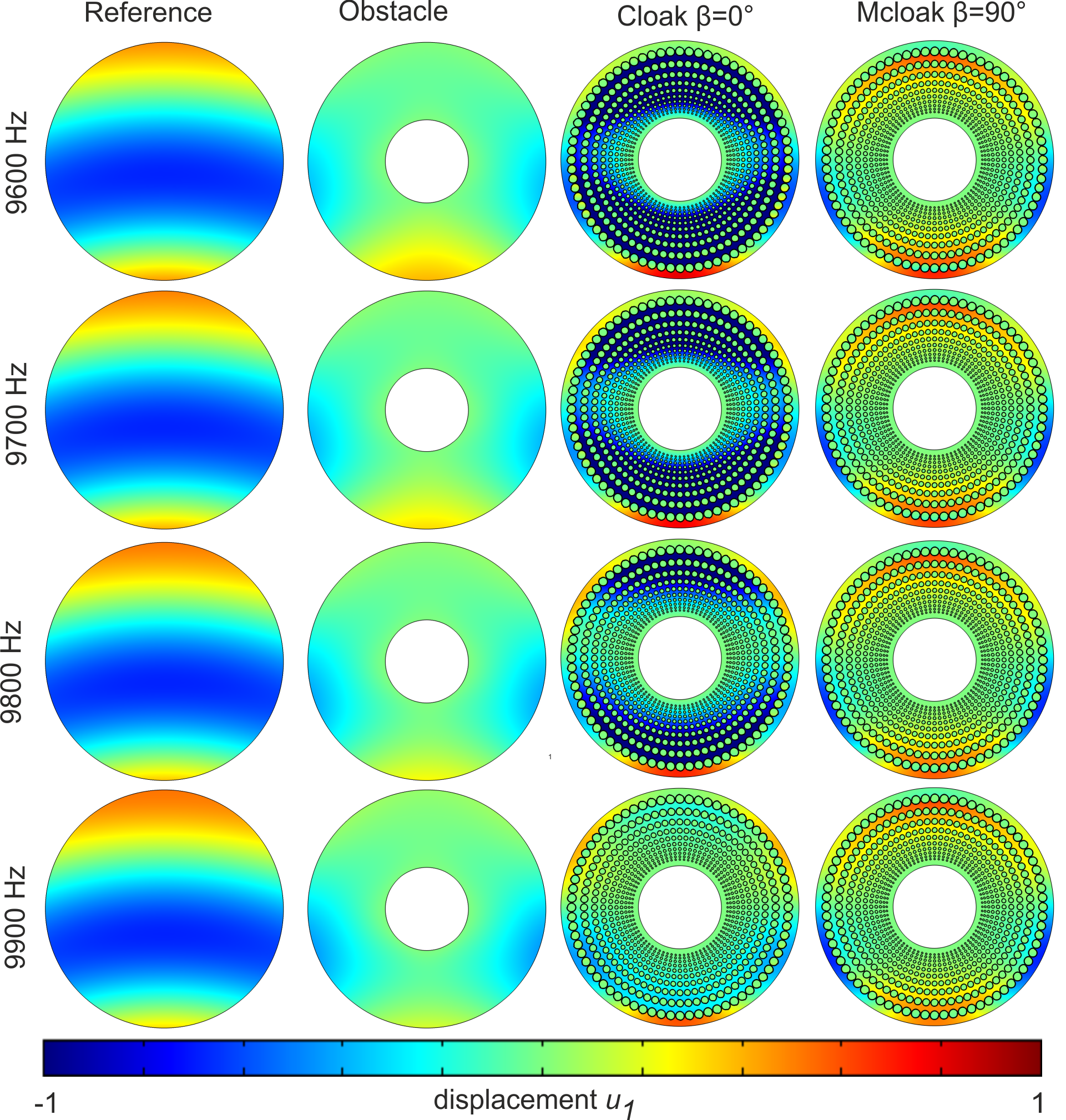}
\caption{Field plots as in Figure 2 but shown only around the cloak's region.}
\label{fig3}}
\end{figure*}

%\begin{figure*}[h!]
%\centering{
%\include-graphics[width=150mm]{Figure4_mod.jpg}
%%\includegraphics[scale=0.5]{u.jpg}
%\caption{Time dependent analysis to illustrate the establishment of the suggested cloaking and the mode conversion induced by the invisibility cloak.}
%\label{fig1}}
%\end{figure*}

%\section{Conclusion}

%%%%%%%%%%%%%%%%%%%%%%% References %%%%%%%%%%%%%%%%%%%%%%%%%

%\vfill\eject

\appendix{Appendix A: On Willis medium with Cosserat coefficients}

We start from a Willis material with elasticity order 4 tensors $\mathbb C^w,$  order 3 tensors $S^w$,  $D^w$ and mass density $\rho^w$ satisfying the Willis Equations
\begin{eqnarray}0&=&\nabla_{\mathbf x}\cdot\Big( \mathbf C^w:\nabla_{\mathbf x}{\mathbf u}+S^w\cdot{\mathbf u}\Big) \nonumber\\
 &&+D^w:\nabla_{\mathbf x}{\mathbf u} +\omega^2\rho^w{\mathbf u}
 \label{willisandre}
 \end{eqnarray}
We apply a transformation ${\mathbf x} \mapsto {\mathbf x'}$  with Jacobian matrix ${\bf J}$ by requiring that the original and transformed displacements be the same. Note that, this is the requirement applied to the transformation scheme that produces Cosserat material, starting from a background material which is isotropic homogeneous \cite{apl2009, diatta-apl2014}. In this Appendix, we pioneer a new scheme consisting on applying this transformation technique to a Willis material, as our background reference material. As one can see below, the resulting transformed material will satisfy the same equations as Willis equations, but will however have elasticity tensor, hereafter denoted by $\mathbb C^{cw}$, which no longer possesses minor symmetries. The resulting tensors $S^{cw}$ and $D^{cw}$ of order 3, still satisfy the identities $S_{ijk}^{cw}=-D_{kij}^{cw}$, but now the coefficients $S_{ijk}^{cw}$  and $ S_{jik}^{cw}$ are generally not equal.

We apply a (general)  transformation ${\bf x}=(x_1,x_2,...)\longmapsto {\bf x'}=\Big(x_1'({\bf x}),x_2'({\bf x}),...\Big)$ mapping $\Omega\subset \mathbb R^n$ to  $\Omega'\subset \mathbb R^n$ and we further impose that the displacement  ${\bf u}'$ in the transformed material filling $\Omega'$, be linearly related to the displacement ${\bf u}$ in the background material (untransformed original material) in $\Omega$, as 
  ${\bf u}'({\bf x'})={\bf J}^{-T} ({\bf x}){\bf u}({\bf x})$ where $J_{ij}=\partial x_i'/\partial x_j$.
We next use the weak formulation of (\ref{willisandre}): for any test function $\phi,$
\begin{eqnarray}
0&=& \displaystyle \int_\Omega \Big[\nabla_{\bf x}\cdot\Big( \mathbf C^w:\nabla_{\mathbf x}{\mathbf u}+S^w\cdot{\mathbf u}\Big) \nonumber \\
&+& D^w:\nabla_{\mathbf x}{\mathbf u}+\rho^w\omega^2 {\bf u}\Big] \cdot \phi ~d{\bf x} \nonumber\\ 
&=&   \displaystyle \int_\Omega \Big[- \Big({\bf C}^w : \nabla_{\bf x}{\bf u}+S^w\cdot{\mathbf u}\Big) :\nabla_{\bf x}{\mathbf \phi} \nonumber \\
&+& \Big(D^w:\nabla_{\mathbf x}{\mathbf u}+\rho^w\omega^2 {\bf u}\Big)\cdot \phi\Big] d{\bf x}
\nonumber\\ 
&=&   \displaystyle \int_{\Omega'} \Big[- \Big({\bf C}^w : \nabla_{\bf x}~{\mathbf J^T}{\bf u'}+S^w\cdot~{\mathbf J^T}{\bf u'}\Big) :\nabla_{\bf x}~{\mathbf J^T}{\mathbf \phi'}
\nonumber\\
&+&~~ \Big( D^w:\nabla_{\mathbf x}{\mathbf J^T}{\bf u'}+\rho^w\omega^2 {\mathbf J^T}{\bf u'}\Big) \cdot {\mathbf J^T} {\mathbf \phi'}\Big] ~\det({\mathbf J}^{-1})~d{\bf x'}
\label{navierweak}
 \end{eqnarray}
%\textcolor{magenta}
{Simply further developing the above we deduce the following}
\begin{eqnarray}0&=&\nabla_{\mathbf x'}.\Big[ \mathbf C^{cw}:\nabla_{\mathbf x'}{\mathbf u}+S^{cw}\cdot{\mathbf u}\Big]\nonumber\\
 &&+D^{cw}:\nabla_{\mathbf x'}{\mathbf u} +\omega^2\rho^{cw}{\mathbf u}
 \end{eqnarray}
with
%\textcolor{magenta}{
\begin{eqnarray}C_{ijkl}^{cw}&=& \frac{1}{\det({\mathbf J})}\displaystyle\sum_{p,q} \frac{\partial x'_i}{\partial x_p} \frac{\partial x'_k}{\partial x_q}  C_{pjql}^{w} ,
 \nonumber\\
 S_{ijk}^{cw}&=&  \frac{1}{\det({\mathbf J})} \displaystyle\sum_{s} \frac{\partial x'_i}{\partial x_s} S_{sjk}^{w}, 
\nonumber\\
 D_{ijk}^{cw}&=& \frac{1}{\det({\mathbf J})} \displaystyle\sum_{s} \frac{\partial x'_j}{\partial x_s} D_{isk}^{w}= - S_{jki}^{cw} , 
\nonumber\\
 {\mathbf \rho}^{cw}&=& \frac{1}{\det({\mathbf J})}{\mathbf \rho}^{w} . 
 \end{eqnarray}

If we apply to the above the radial transformation $(r,\theta) \mapsto (r',\theta'):= (\frac{r_2-r_1}{r_2}r+r_1, ~\theta)$ in polar coordinates, then the only non vanishing coefficients
of the Cosserat-Willis tensor ${\bf C}^{cw}$ are 
\begin{eqnarray}C_{r'r'r'r'}^{cw}&=& \frac{r'-r_1}{r'}  C_{r'r'r'r'}^{w} , ~C_{r'r'\theta\theta}^{cw}= C_{r'r'\theta\theta}^{w} , 
\nonumber\\ C_{r'\theta r'\theta}^{cw}&=& \frac{r'-r_1}{r'}  C_{r'\theta r'\theta}^{w} , ~C_{r'\theta\theta r'}^{cw}= C_{r'\theta\theta r'}^{w} ,\nonumber\\
 C_{\theta r' r'\theta}^{cw}&=&  C_{\theta r' r'\theta}^{w} , ~C_{\theta r'\theta r'}^{cw}= \frac{r'}{r'-r1}  C_{\theta r'\theta r'}^{w} ,
\nonumber\\
~C_{\theta\theta r' r'}^{cw}&= &  C_{\theta\theta r' r'}^{w} , ~C_{\theta\theta\theta\theta}^{cw}= \frac{r'}{r'-r_1}  C_{\theta\theta\theta\theta}^{w}.
 \nonumber\\
 S_{r'jk}^{cw}&=& \frac{r_2}{r_2-r_1} \frac{r'-r_1}{r'} S_{r'jk}^{w} =-D_{k r'j}^{cw} , ~j,k= r',\theta,
\nonumber\\
 S_{\theta jk}^{cw}&=& \frac{r_2}{r_2-r_1}  S_{\theta jk}^{w}=-D_{k\theta j}^{cw}, ~j,k=r',\theta'.
% \nonumber\\
% S_{ijk}^{cw}&=&  \frac{1}{\det(J_{\mathbf y \mathbf x})} \displaystyle\sum_{s} \frac{\partial y_i}{\partial x_s} S_{sjk}^{w} = - D_{kij}^{cw}, 
\label{tensorandre}
 \end{eqnarray}
 and the Cosserat-Willis density is
 \begin{equation}
 {\mathbf \rho}^{cw}=\frac {r_2^2}{(r_2-r_1)^2}\frac{r'-r_1}{r'}{\mathbf \rho}^{w} .
 \label{densityandre} 
 \end{equation}


\begin{thebibliography}{99}

\bibitem{Milton}
G.W. Milton, M. Briane and J.R. Willis,
On cloaking for elasticity and physical equations with a transformation invariant form,
New Journal of Physics {\bf 8}, (2006).

\bibitem{apl2009}
M. Brun, S. Guenneau and A.B. Movchan,
Achieving control of in-plane elastic waves,
Applied Physics Letters {\bf 94}, 061903 (2009).

\bibitem{diatta-apl2014}
A. Diatta and S. Guenneau, Controlling solid elastic waves with spherical cloaks. Applied Physics Letters {\bf 105}, 021901 (2014).

\bibitem{Norris}
A. Norris and A.L. Shuvalov, Elastic cloaking theory, Wave Motion {\bf 48}(6), 525-538 (2011).

\bibitem{Mccall}
M. McCall et al.,
Roadmap on transformation optics,
Journal of Optics {\bf 20}(6), 63001 (2018).

\bibitem{Willis}
J.R. Willis, Variational principles for dynamic problems for inhomogeneous elastic media, Wave Motion {\bf 3}, 1-11 (1981).

\bibitem{pendry}
J.B. Pendry,
A new route to negative refraction,
Science {\bf 306}, 1353-1355 (2004).

\bibitem{Frenzel}
T. Frenzel, M. Kadic, M. Wegener,
Three-dimensional mechanical metamaterials with a twist,
Science {\bf 358}(6366), 1072-1074 (2017).

\bibitem{fernandez2019}
I. Fernandez-Corbaton, C. Rockstuhl, P. Ziemke, P. Gumbsch, A. Albiez, R. Schwaiger, T. Frenzel, M. Kadic, M. Wegener,
New twists of 3D chiral metamaterials,
Advanced Materials, 1807742 (2019).

\bibitem{kadic2019}
M. Kadic, G.W. Milton, M. van Hecke, M. Wegener,
3D metamaterials,
Nature Reviews Physics, 1 (2019).

\bibitem{kadic2019b}
M. Kadic, A. Diatta, T. Frenzel, S. Guenneau, M. Wegener,
Static chiral Willis continuum mechanics for three-dimensional chiral mechanical metamaterials,
Physical Review B {\bf 99}, 214101 (2019).

\bibitem{Quan}
L. Quan, Y. Radi, D. L. Sounas, and A. Alu,
Maximum Willis Coupling in Acoustic Scatterers,
Physical Review Letters {\bf 120}(25), 254301 (2018).

\bibitem{Melnikov}
A. Melnikov, Y.K. Chiang, L. Quan, S. Oberst, A. Alu, S. Marburg, and D. Powell, 
Acoustic meta-atom with experimentally verified maximum Willis coupling,
Nature Communications {\bf 10}, 3148 (2019).

\bibitem{Russell}
P.S.J. Russell, E. Marin, A. Diez, S. Guenneau and A.B. Movchan,
Sonic band gaps in PCF preforms: enhancing the interaction of sound and light,
Optics Express {\bf 11}(20), 2555-2560 (2003).

%\bibitem{morse}
%P.M. Morse, H. Feshbach, Methods of theoretical physics, pp. 52-54. McGraw-Hill, 1953.

%\bibitem{Andrianov}
%I.V. Andrianov,  V.I. Bolshakov, V.V. Danishevs’kyy, D. Weichert,
%Higher order asymptotic homogenization and wave propagation in periodic
%composite materials. Proceedings of the Royal Society A: Mathematical,
%Physical and Engineering Sciences {\bf 464}, 1181-1201 (2008).

%\bibitem{Auriault}
%J.L. Auriault, C. Boutin, Long wavelength inner-resonance cut-off
%frequencies in elastic composite materials. International Journal of Solids
%and Structures {\bf 49}, 3269-3281 (2012).

\bibitem{Nassar}
H. Nassar, Q.C. He, and N. Auffray,
A generalized theory of elastodynamic homogenization for periodic media
International Journal of Solids and Structures {\bf 84}, 139-146 (2016).

\bibitem{Kadic}
M. Kadic, A. Diatta, T. Frenzel, S. Guenneau, and 
M. Wegener,
Static chiral Willis continuum mechanics for three-dimensional chiral mechanical metamaterials,
Physical Review B {\bf 99}, 214101 (2019).

\bibitem{physicab}
S. Guenneau, F. Zolla,
Homogenization of 3D finite chiral photonic crystals,
Physica B {\bf 394}, 145-147 (2007).

\bibitem{Achaoui}
Y. Achaoui, A. Diatta and S. Guenneau,
Steering in-plane shear waves with inertial resonators in platonic crystals,
Applied Physics Letters {\bf 106}(22), 223502 (2015).

\bibitem{buckmann15}
T. B\" uckmann, M. Kadic, R. Schittny and M. Wegener,
Mechanical cloak design by direct lattice transformation,
Proceedings of the National Academy of Sciences {\bf 112}(16), 4930-4934 (2015).

\bibitem{kadic2019c}
M. Kadic, M. Wegener, A. Nicolet, F. Zolla, S. Guenneau and A. Diatta,
Dynamic behavior of mechanical cloaks designed by direct lattice transformation,
Wave Motion (in press)

\bibitem{nassar18}
H. Nassar, Y.Y. Chen and G.L. Huang,
A degenerate polar lattice for cloaking in full two-dimensional elastodynamics and statics
Proceedings of the Royal Society A {\bf 474}(2219), 20180523 (2018).

\end{thebibliography}
\end{document}